\documentclass{webofc}

\usepackage[varg]{txfonts}   
\usepackage{hyperref}
\usepackage{url}
\hypersetup{colorlinks=true,citecolor=blue,urlcolor=blue,linkcolor=blue}
\begin{document}
\title{Simulating stochastic fluid dynamics}
%
%

\author{
 \firstname{Chandrodoy} \lastname{Chattopadhyay}\inst{1}
 \and
 \firstname{Josh} \lastname{Ott}\inst{2}
 \and
 \firstname{Thomas} \lastname{Sch\"afer}\inst{1} 
 \fnsep\thanks{\email{tmschaef@ncsu.edu}}
 \and
 \firstname{Vladimir} \lastname{Skokov}\inst{1}
}

\institute{Department of Physics, North Carolina State
University, Raleigh, NC 27695
\and
Center for Theoretical Physics -- a Leinweber Institute, Massachusetts Institute of Technology, Cambridge, MA 02139
}

\abstract{We present simulations of the real time dynamics
of a fluid in the vicinity of a critical endpoint in the 
phase diagram. The relevant hydrodynamic theory is known 
as model H, and it is expected to describe the long-distance
dynamics of QCD matter in the vicinity of a possible critical
endpoint in the QCD phase diagram.
}
\hfill MIT-CTP/5914
\maketitle

\section{Introduction}
\label{sec-intro}

  One of the goals of the heavy ion program at RHIC, the LHC, and
future facilities at lower energy is to find direct evidence for 
a phase transition between a quark-gluon phase and a hadronic phase 
of QCD. Two ideas that have been discussed are to observe 
fluctuations signatures associated with the chiral crossover 
transition at zero chemical potential, or to observe critical 
fluctuations related to a possible endpoint of a first order 
transition at non-zero chemical potential. 

 The critical dynamics in the vicinity of a phase transition
is described by fluid dynamic theories that contain stochastic
terms in addition to the well-known gradient expansion. These
stochastic terms ensure that fluctuation-dissipation relations 
are satisfied, in particular in a regime where fluctuations are 
not small. A theory that describes the universal critical 
dynamics near an endpoint in the universality class of the 
liquid-gas transition is model H \cite{Hohenberg:1977ym}. This 
theory is expected to describe the dynamics of a QCD fluid 
in the vicinity of a possible endpoint of a first order 
transition in the phase diagram \cite{Son:2004iv}.

 There is a significant amount of literature on the dynamics
of fluctuations in QCD \cite{Basar:2024srd}, but model H 
has never been numerically simulated. This is the case because
there are a number of difficulties that need to be overcome. 
Stochastic fluid dynamic models have fluctuations on all 
scales, ranging from the size of the system to the microscopic
scale at which the theory is discretized. These fluctuations
lead to ultra-violet divergences that have to be renormalized, 
and the regularization and renormalization procedure has to 
be compatible with fluctuation-dissipation relations and 
stability requirements. In the following we briefly explain
a new approach to this problem, and summarize the results 
obtained in our recent work \cite{Chattopadhyay:2024jlh,
Chattopadhyay:2024bcv}.

\section{Model H}
\label{sec-mod-H}

 Model H describes the interaction of an order parameter density 
$\phi$ with the momentum density $\vec\pi$ of the fluid. This 
equations of motion are given by  \cite{Hohenberg:1977ym}
\begin{align}
\label{modH_1}
\partial_t\phi &=   \Gamma\,\nabla^2 
        \left(\frac{\delta{\cal H}}{\delta \phi}\right)
- \left(\nabla_i\phi\right) \frac{\delta{\cal H}}{\delta \pi_i^T}
      + \zeta , \\
\label{modH_2}
\partial_t \pi^T_i &= \eta \nabla^2 
    \left(\frac{\delta{\cal H}}{\delta \pi^T_i}\right)
    +  P^T_{ij} \left[\left(\nabla_j\phi\right) \frac{\delta{\cal H}}
    {\delta\phi}  \right] 
    -  P^T_{ij} \left[ 
        \nabla_k\left( \pi^{T}_j \frac{\delta{\cal H}}{\delta
       \pi^T_k}\right) \right] + \xi_i\, . 
\end{align}
For the purpose of describing the dynamics near a liquid-gas endpoint 
we can take $\phi$ to be the specific entropy $s/n$ of the fluid
\cite{Akamatsu:2018vjr}. $\Gamma$ is the thermal conductivity
and $\eta$ is the shear viscosity. The transverse projection 
operator is given by $P^T_{ij} = \delta_{ij}- \nabla_i\nabla_j/
\nabla^2$ and the transverse momentum density is $\pi^T_i=
P^T_{ij}\pi_j$. The effective Hamiltonian is given by 
\begin{align}
\label{H_Ising}
    {\cal H}  = \int d^dx \left[ 
      \frac{1}{2\rho} ( \pi_i^T)^2
    +  \frac{1}{2} (\nabla \phi)^2  
    +  \frac{1}{2} m^2 \phi^2
    +   \frac{1}{4} \lambda  \phi^4  
    - h \phi\right] \, ,
\end{align}
where $\rho$ is the mass density, $m$ is the inverse
correlation length, $\lambda$ is a non-linear self-coupling, 
and $h$ is an external field. At the mean field level this 
Hamiltonian has a critical point at $h=0$ and $m^2=0$. When 
fluctuations are included the critical value of $m^2$ at 
$h=0$ has to be determined numerically. In order to describe
a critical endpoint in the QCD phase diagram the parameters 
$m^2$, $h$, and $\lambda$ will have to be mapped onto the 
chemical potential-temperature $(\mu,T)$ plane in QCD, along 
the lines described in \cite{Parotto:2018pwx}. The noise fields
$\zeta$ and $\xi_i$ are random variables constrained by
fluctuation-dissipation relations. The correlation functions 
of the noise fields are given by  
\begin{align}
    \langle \zeta (t, \vec{x}) \zeta (t', \vec{x}') \rangle &= 
    -2 T\, \Gamma\, \nabla^2 \delta(\vec{x}-\vec{x}')\delta(t-t')\, ,
\label{noise-phi}  \\  
  \langle \xi_i (t, \vec{x}) \xi_j (t', \vec{x}') \rangle &= 
    -2 T\, \eta\, P^T_{ij} \nabla^2
      \delta(\vec{x}-\vec{x}')\delta(t-t')\, . 
\label{noise-pi} 
\end{align}

\begin{figure}[t]
\begin{center}
\includegraphics[width=5.5cm]{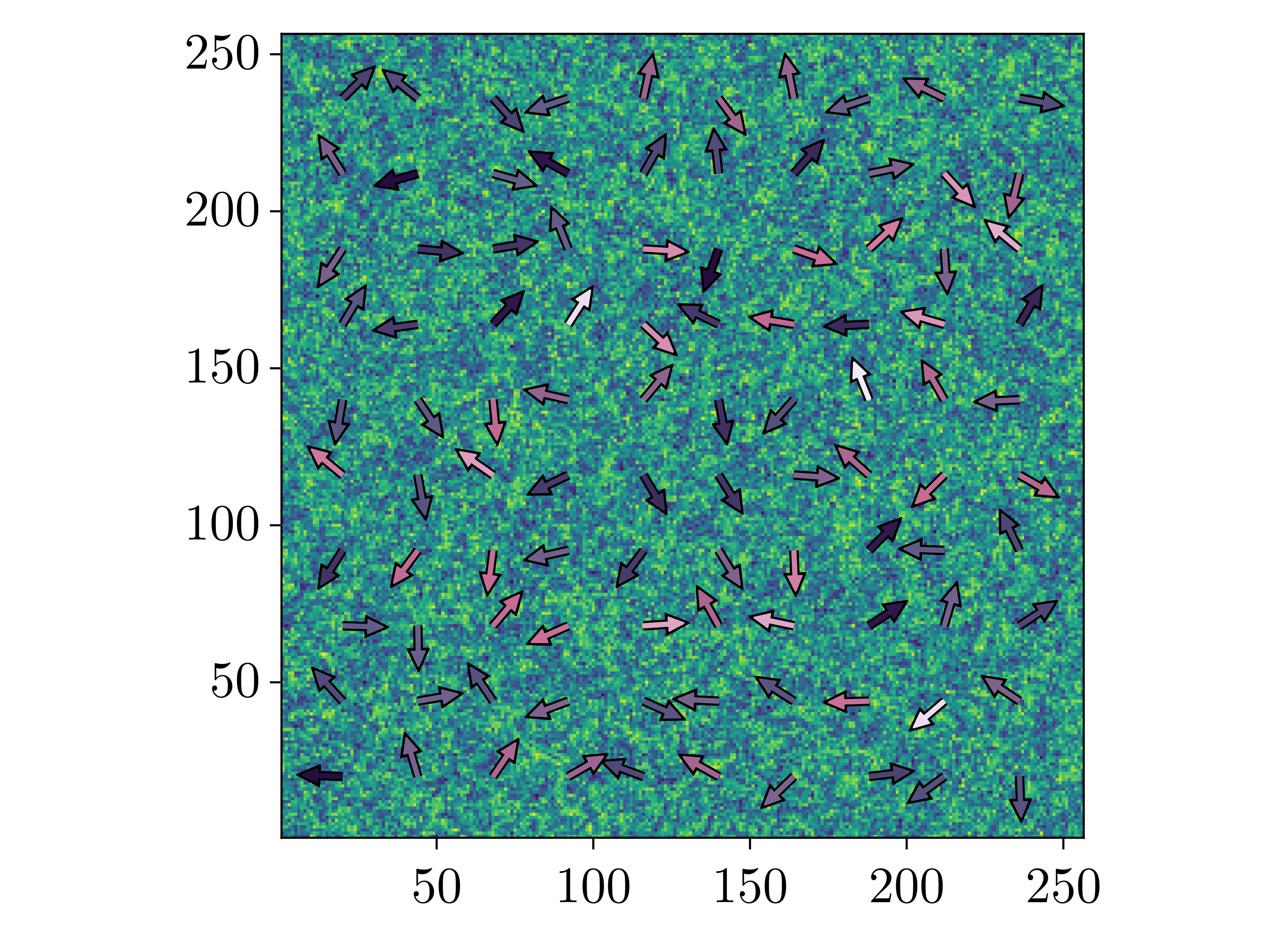}
\hspace*{0.4cm}
\includegraphics[width=5.5cm]{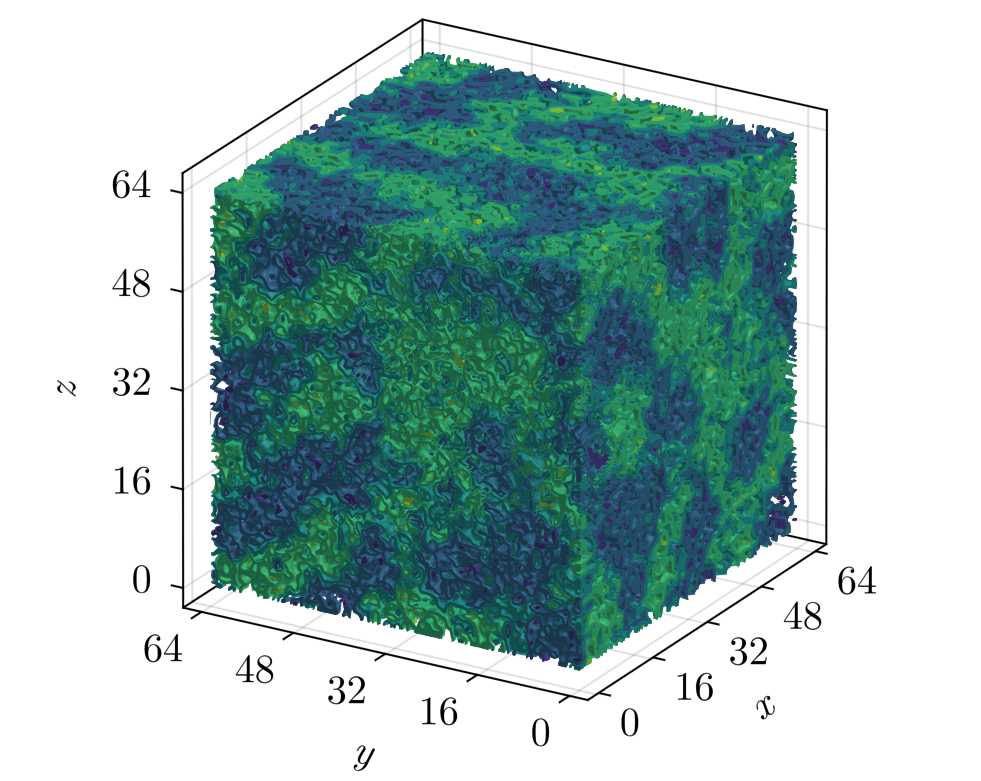}
\end{center}
\caption{Left panel: Order parameter $\phi$ (color coded) and 
fluid momentum $\vec\pi$ (arrows) configuration in a stochastic
two-dimensional fluid. Right panel: Order parameter configuration
in a three dimensional fluid. 
}
\label{fig-1}       
\end{figure}

\section{Numerical Methods}
\label{sec-num}

  The main idea underlying the numerical algorithm is to 
combine the dissipative and stochastic updates into a 
single Metropolis update. The update for the order parameter 
field is 
\begin{align}
\begin{array}{rcl}
\phi^{\it trial}(\vec{x},t+\Delta t) & = & \phi(\vec{x},t)
   + q_\mu\, , \\
\phi^{\it trial}(\vec{x}+\hat\mu,t+\Delta t) &=&
   \phi(\vec{x}+\hat{\mu},t)  - q_\mu\,  ,
   \end{array}
   \hspace*{1cm}
 q_\mu = \sqrt{2\Gamma T(\Delta t)}\, \xi  \, ,
\label{phi-stoch}
\end{align}
where $\xi$ is a Gaussian random variable with unit variance
and $\hat{\mu}$ is an elementary lattice vector in the direction
$\mu=1,\ldots,d$. The update is accepted with probability
${\it min}(1,e^{-\Delta{\cal H}/T})$. Note that this algorithm is
automatically conserving. This algorithm
is based on the observation that the average update $\langle            
[\phi(\vec{x},t+\Delta t) - \phi(\vec{x},t)]\rangle$ realizes
the diffusion equation, and the second moment $\langle [\phi            
(\vec{x},t+\Delta t)-\phi(\vec{x},t)]^2\rangle$ reproduces the
noise term, see \cite{Florio:2021jlx}. We follow the same procedure 
for $\vec\pi$. We perform a conserving Metropolis update for 
all components of the momentum density, and then apply a transverse
projection operator. 

 In principle, the advection step can be discretized using standard
methods in computational fluid dynamics. However, some attention has
to be paid to the fact that the fields exhibit large fluctuations 
on the scale of the lattice spacing, even as the lattice spacing 
is taken to zero. This means that we may encounter large violations
of conservation laws that follow from the ideal equations of motion
in the continuum limit. In practice we have used a discretization method
that employs ``skew-symmetrized derivatives'' which has been 
developed for the purpose of simulating incompressible turbulence
\cite{Morinishi:1998}.
 
\begin{figure}[t]
\centering
\includegraphics[width=6.cm,clip]{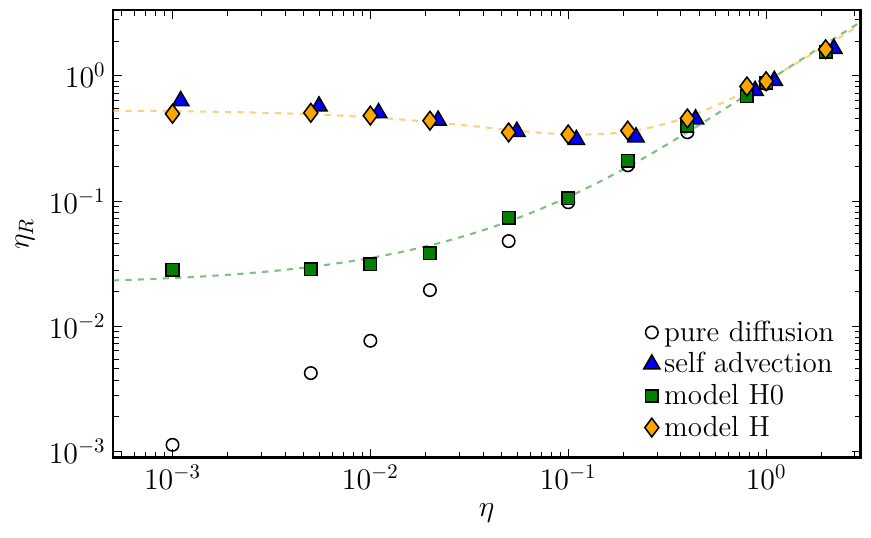}
\includegraphics[width=6.cm,clip]{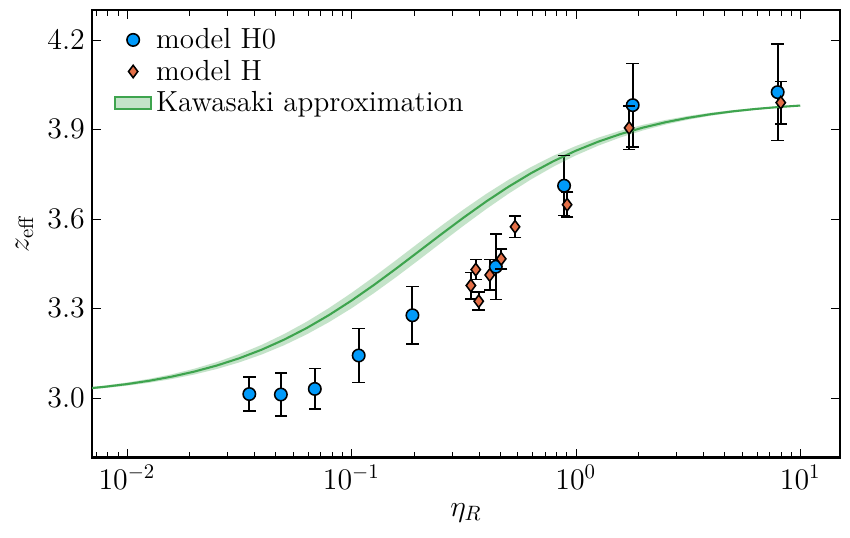}
\caption{
\label{fig-2}       
Left panel: Physical shear viscosity as a function of the bare 
viscosity in a non-critical fluid. Several truncations are 
shown: Model H0 (no self-advection), self-advection only, and
pure diffusion (no interactions). Right panel: Dynamical 
critical exponent $z$ extracted from a scaling analysis of 
the order parameter correlation function in a finite volume 
as a function of the physical viscosity. }
\end{figure}

\section{Results and outlook}
\label{sec-res}

 We have performed simulations of critical and non-critical fluid 
in box of volume $L^3$, see Fig.~\ref{fig-1}. We have verified that 
$m^2$ can be tuned to a critical point at which the simulation 
reproduces the static critical exponents of the 3d Ising Model.
The simplest dynamic phenomenon that we have studied is the 
renormalization of the shear viscosity $\eta$ of a non-critical fluid, 
see the left panel of Fig.~\ref{fig-2}. Perturbative calculations
suggest that there is a UV sensitive renormalization of $\eta$
that is inversely proportional to the bare value of $\eta$, an effect 
sometimes called the ``stickiness of sound'' \cite{Kovtun:2011np}.
We clearly observe this effect, and we find it to be dominated 
by the self-advection of the momentum density. This effect is 
included in the full model H simulation, but dropped in the truncation
denoted as model H0. 

 We have also measured the dynamic critical exponent $z$. This quantity
is a measure of critical slowing-down, the increase of the relaxation
time $\tau\sim \xi^z$ as the correlation length $\xi$ increases. We
have measured this effect directly at the critical point by varying 
the system size $L$, which limits the correlation length in a 
finite volume. We observe a value $z\simeq 3.01$ which is clearly
different from mean-field behavior, but consistent with predictions
from the epsilon-expansion. We also observe that in any finite 
system there is a crossover from model B like behavior $z\simeq 4$
to model H scaling $z\simeq 3$ as the viscosity is reduced. Finally,
we observe evidence for universality: The scaling exponent in model H
and model H0 is the same when plotted as a function of the physical 
viscosity (see Fig.~\ref{fig-2} right panel). 

 We have demonstrated that we can also measure more complicated 
observables, such as the relaxation rate of higher moments of the 
order parameter. Future work will focus on trying to model realistic
systems, by embedding the framework discussed here in an expanding 
relativistic fluid. 

 Acknowledgements: We acknowledge support by the DOE Office of 
Science under contracts DE-FG02-03ER41260 (TS) and DE-SC0020081 (VS).

%
%
\bibliography{bib} 
%
%
%

\end{document}